\documentclass{article}
\usepackage{spconf,amsmath,graphicx}
\usepackage{amssymb}
\usepackage{bbding}
\usepackage{booktabs}
\usepackage[hidelinks]{hyperref}

\title{Multi-Task Deep Residual Echo Suppression with Echo-aware Loss}
\name{Shimin Zhang$^{1}$, Ziteng Wang, Jiayao Sun$^{1}$, Yihui Fu$^{1}$, Biao Tian, Qiang Fu, Lei Xie$^{1*}$
\thanks{$^*$: Corresponding author.}
}
\address{
	$^{1}$Audio, Speech and Language Processing Group (ASLP@NPU), School of Computer Science\\
	Northwestern Polytechnical University, Xi'an, China\\
\texttt{shmzhang@npu-aslp.org, lxie@nwpu.edu.cn}
}
\begin{document}
\ninept
\maketitle
\begin{abstract}

\end{abstract}

This paper introduces the NWPU Team's entry to the ICASSP 2022 AEC Challenge. We take a hybrid approach that cascades a linear AEC with a neural post-filter. The former is used to deal with the linear echo components while the latter suppresses the residual non-linear echo components. We use gated convolutional F-T-LSTM neural network (GFTNN) as the backbone and shape the post-filter by a multi-task learning (\textit{MTL}) framework, where a voice activity detection (VAD) module is adopted as an auxiliary task along with echo suppression, with the aim to avoid over suppression that may cause speech distortion. Moreover, we adopt an \textit{echo-aware} loss function, where the mean square error (MSE) loss can be optimized particularly for every time-frequency bin (TF-bin) according to the signal-to-echo ratio (SER), leading to further suppression on the echo. Extensive ablation study shows that the time
delay estimation (TDE) module in neural post-filter leads to better perceptual quality, and an adaptive filter with better convergence will bring consistent performance gain for the post-filter. Besides, we find that using the linear echo as the input of our neural post-filter is a better choice than using the reference signal directly. In the ICASSP 2022 AEC-Challenge, our approach has ranked the 1st place on word accuracy (WAcc) (0.817) and the 3rd place on both mean opinion score (MOS) (4.502) and the final score (0.864). 

\section{Introduction}
\label{sec:intro}
Annoying acoustic echo arises when microphone and loudspeaker are coupled in a real-time communication (RTC) system such that the microphone picks up the desired near-end speech signal plus the unwanted loudspeaker signal.
Acoustic echo has become one of the major sources of poor speech quality ratings in RTC~\cite{sridhar2020icassp}. As an indispensable module in RTC, \textit{acoustic echo cancellation} (AEC) aims to remove such unpleasant acoustic echo from the near-end microphone signal while minimizing the distortion of the near-end speaker's speech. More challengingly, a complete AEC system may need to consider noise suppression~\cite{zhang2018deep} as well because noise is inevitable in actual acoustic environments. Moreover, AEC is also essential for downstream tasks such as speech recognition~\cite{o2021conformer}.

%many scenarios include speech recognition~\cite{cutler2022AEC}

Digital signal processing (DSP) based linear echo cancellation has been adopted in RTC for a long time, which works by estimating near-end speech~\cite{ferrara1980fast} or acoustic echo path~\cite{enzner2006frequency} with an adaptive filter. Such algorithms usually give an estimate of the near-end speech and an estimate of the linear echo obtained by cancellation. A post-filter algorithm~\cite{hansler2005acoustic} is usually used to further remove the residual echo that cannot be completely removed by the linear AEC algorithm. However, these methods are still not quite effective for echo suppression especially in the presence of noise interference.

Recent advances in deep learning have shown great potential in AEC due to the strong non-linear modeling ability of deep neural networks (DNN). Some neural AEC approaches tried to use a neural network directly without any conventional DSP module. Specifically, Zhang and Wang~\cite{zhang2018deep} formulated AEC as a supervised speech separation problem, where a bidirectional long-short term memory (LSTM) network was adopted to predict a magnitude mask of the microphone signal.  West-hausen et al.~\cite{westhausen2020acoustic} extended dual-signal transformation LSTM network (DTLN)~\cite{Westhausen2020} by adopting the far-end signal as additional information. Zhang et al.~\cite{zhang2021ft} extended deep complex convolution recurrent network (DCCRN)~\cite{hu2020dccrn}, which was originally designed for speech enhancement with superior performance, with frequency-time LSTM (F-T-LSTM) network~\cite{li2015lstm} to better learn the relationship between frequency bands for effectively suppressing echo.
In contrast, other methods effectively combine DSP with a neural network without losing the advantages of traditional signal processing in handling the linear component of mixed signal. As a typical approach, Wang et al.~\cite{wang2021weighted} cascaded an adaptive filter based linear echo canceller with a deep feedforward sequential memory network (DFSMN)~\cite{zhang2018dfsmn} based post-filter. Peng et al.~\cite{peng2021acoustic} adopted multiple filters for simultaneous linear echo cancellation and time delay estimation (TDE), cascaded further with a gated complex convolutional
recurrent neural network (GCCRN) post-filter. The two approaches have shown superior performance in the recent AEC-Challenge series~\cite{sridhar2020icassp, cutler2021interspeech}.

Recently, processing of super-wideband and full-band speech signal has become more and more popular with the explosion of real-time communication and online collaboration. However, modeling more frequency bands and particularly high frequency components of speech signal is more challenging. The latest ICASSP 2022 AEC Challenge~\cite{cutler2022AEC} has particularly focused on AEC for \textit{full-band} speech. More challengingly, the performance of downstream speech recognition is also considered as a challenge metric. In other words, the AEC performance is measured by both human and machine auditory perception, i.e., human subjective listening score and machine speech recognition accuracy.

In this paper, we introduce our NWPU Team's entry to the ICASSP 2022 AEC Challenge. We submit a \textit{hybrid} approach that combines a linear AEC with a neural post-filter. Specifically, inspired by our recent work~\cite{zhang2021ft}, we propose a gated convolutional F-T-LSTM neural network (GFTNN) as the post-filter.
More importantly, we shape our neural post-filter by a multi-task learning (\textit{MTL}) framework, where a voice activity detection (VAD) module is adopted as an auxiliary task along with echo suppression. The MTL framework is designed to avoid over suppression and benefit speech recognition accuracy. Moreover, we adopt an \textit{echo-aware} loss function, where the mean square error (MSE) loss can be optimized particularly for every time-frequency bin (TF-bin) according to the signal-to-echo ratio (SER), leading to further suppression of the echo. Through an extensive ablation study, we also find that the TDE module leads to better perceptual quality, and an adaptive filter with better convergence will bring consistent performance gain for the post-filter. Besides, using the linear echo as the input of our neural post-filter is a better choice than using the reference signal directly. According to the official results of the challenge, 
our approach has ranked the 1st place on word accuracy (WAcc) (0.817) and the 3rd place on both mean opinion score (MOS) (4.502) and the final score (0.864).

%  \vspace{-5pt}
\section{Proposed Method}
\label{sec:method}

\begin{figure}
\centering
    % \vspace{20pt}
    \includegraphics[width=0.5\textwidth]{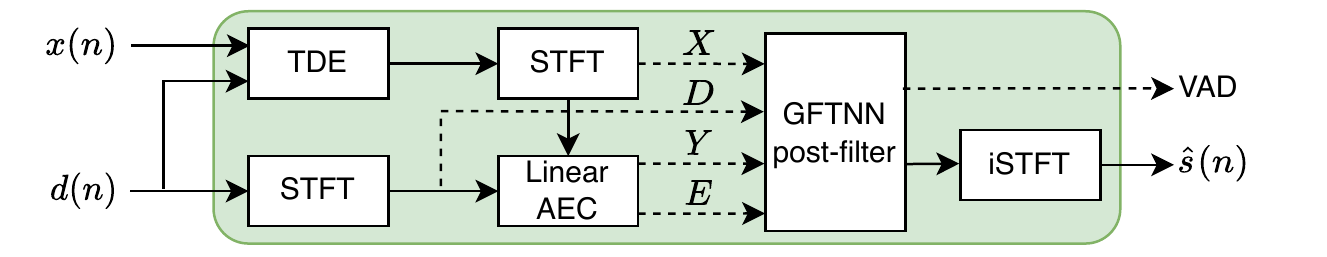}
     \vspace{-10pt}
    \caption{Diagram of our proposed AEC system.}
    \label{fig:aec_flow}
    \vspace{-10pt}
\end{figure}

%  \vspace{-5pt}
\subsection{Problem formulation}
\label{sec:formula}

We illustrate the main architecture of the proposed AEC algorithm in Fig.~\ref{fig:aec_flow}. The microphone signal $d(n)$ consists of near-end speech $s(n)$, acoustic echo $z(n)$ and background noise $v(n)$:
{
\setlength\abovedisplayskip{2.0pt}
\setlength\belowdisplayskip{2.0pt}
\begin{equation}
\footnotesize
    d(n) = s(n) + z(n) + v(n)
\end{equation}
}
where $n$ refers to the time sample index. Here $z(n)$ is obtained by the far-end signal $x(n)$ convolved by echo path, with potential nonlinear distortions caused by loudspeakers. The AEC task aims to separate $s(n)$ apart from $d(n)$, on the premise that $x(n)$ is known. The error signal $e(n)$ and linear echo $y(n)$ are generated using $x(n)$ and $d(n)$ by adaptive filter. $D, E, X$ and $Y$ are the frequency domain representation of $d, e, x$ and $y$, respectively. Note that the dotted lines in Fig.~\ref{fig:aec_flow} are optional according to different configurations, which will be compared in Section~\ref{sec:exp}. Specifically, the VAD module attached with the post-filter is used as the auxiliary task.

%  \vspace{-5pt}
\subsection{Split and synthesize}

To confine our model with reasonable size and complexity, here we use 3-band finite impulse response (FIR) filter-bank with discrete cosine transform (DCT) modulation~\cite{harris2021multirate} to decompose the full-band (48k Hz) signal to subbands and only process the wide-band (16k Hz) signals, and the full-band signal finally is synthesized by average gain control.

As shown in Fig.~\ref{fig:system}(a), for full-band signal $d_{full}$ and $x_{full}$, we use band-pass filters to obtain the wide-band signal $d$ and $x$, where $d_h$ denotes the remaining high-frequency bands of signal (8 to 16k Hz and 16 to 24k Hz). The average gain each frame for $d_h$ is calculated as follows:
{
\setlength\abovedisplayskip{2.0pt}
\setlength\belowdisplayskip{2.0pt}
\begin{equation}
\footnotesize
g(t) = \min(\frac{\sum_{a}^{b}\hat S{(t,f)}}{\sum_{a}^{b}D{(t,f)}}, \frac{\sum_{c}^{d}\hat S{(t,f)}}{\sum_{c}^{d}D{(t,f)}})
\end{equation}
}
where $t$ is frame index and $f$ denotes frequency bin index, $\{a=11, b=81\}$ and $\{c=121, d=161\}$ cover frequency range for 0.5 to 4k Hz and 6 to 8k Hz, respectively. 

In the pipeline of the AEC system in Fig.~\ref{fig:aec_flow}, we implement TDE module using Generalized Cross Correlation with PHAse Transform (GCC-PHAT)~\cite{knapp1976generalized} and the linear AEC using MDF~\cite{soo1990multidelay} and wRLS~\cite{wang2021weighted}. Specifically, we make comparison on different adaptive filters which will be explained in detail in Section~\ref{sec:exp} later. The process of subband synthesis is as follows:
{
\setlength\abovedisplayskip{2.0pt}
\setlength\belowdisplayskip{2.0pt}
\begin{equation}
\footnotesize
    \hat{S}_{full}(t,f) = \text{SYN}\big(\hat S{(t,f)}, g(t) \cdot D_{h}(t,f)\big)
\end{equation}
}
where $\hat{S}, \hat{S}_{full}$ and $D_h$ are the frequency domain representation of $\hat s, \hat s_{full}$ and $d_h$. $\text{SYN}$ represents commonly used subband synthesis method in RTC scenarios~\cite{harris2021multirate}.

%  \vspace{-5pt}
\subsection{GFTNN post-filter}
\label{ssec:arc}
% architecture

\begin{figure*}[htb]
\vspace{-10pt}
\centering
    \includegraphics[width=0.91\textwidth]{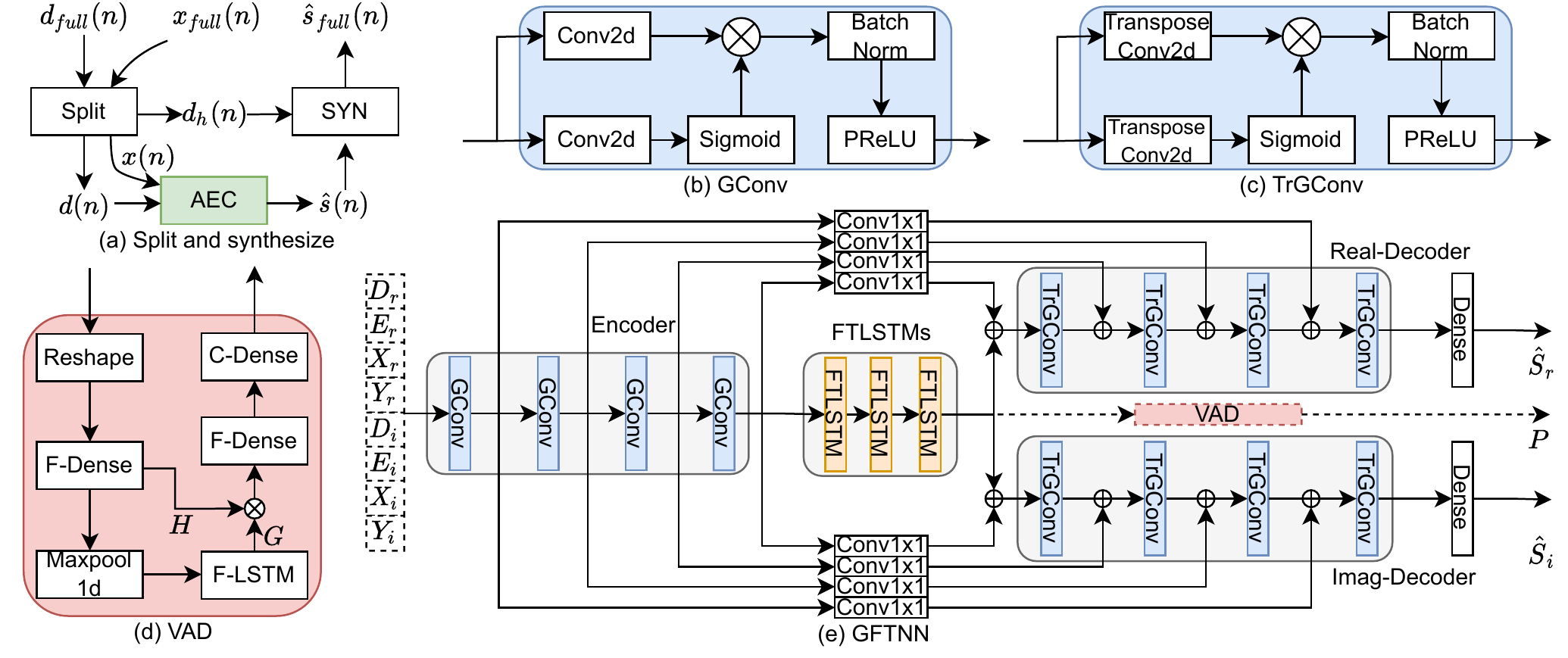}
    \vspace{-10pt}
    \caption{Full-band signal processing and GFTNN post-filter. (b) GConv: Gated conv-2d. (c) TrGConv: Transpose gated conv-2d.}
    \label{fig:system}
     \vspace{-7pt}
\end{figure*}

Fig.~\ref{fig:system}(b), (c), (d) and (e) show the four sub-modules of the proposed GFTNN-based post-filter, namely GConv, TrGConv, VAD and FTLSTM. $*_r$/$*_i$ represents the real/imaginary part of a certain signal in frequency domain. The dotted box of the input feature indicates that a certain signal may not be used. In this paper, we explore three types of combinations, namely $DX$, $EX$ and $DEY$.

Taking $DEY$ as an example, the input feature ${w} \in \mathbb{R}^{3 \times n}$, where $3$ denotes three signals -- $d(n), e(n)$ stacks with $y(n)$. Performing short-time Fourier transform (STFT) on the input feature ${w}$, we obtain the complex spectra ${W} = {W_r} + j{W_i}$, where ${W} \in \mathbb{R}^{6 \times T \times F}$. $T$ denotes the frame number and $F$ denotes the frequency bins.  We use the method in~\cite{peng2021acoustic} to compress ${w}$ and decompress them after the last layer of the GFTNN post-filter.

The GFTNN encoder consists of 4 GConv layers with same input and output channel for each Conv2d layer, excepting for the first GConv layer which needs to be selected according to the input signal (6 in the given example). $1\times1$ convolutional layers are used to connect shallow and deep feature representation, which proves to be beneficial according to~\cite{silva2021acoustic}. The FTLSTM layers (FTLSTMs) are defined in~\cite{zhang2021ft}. The output tensor dimension of both the encoder and the FTLSTMs is $\mathbb{R}^{C \times T \times F}$, where $C$ denotes the channel dimension.

The Real/Imag-Decoder consists of 4 TrGConv layers with same input and output channel for each Transpose-Conv2d layer, except that the output channel of the last TrGConv layer needs to be 1, which is used to estimate the real/imaginary part of the clean near-end spectra. $\oplus$ indicates the $\text{conv}1\times1$ output concatenated with the previous TrGConv layer output.

\begin{table}[]
\footnotesize
\caption{Configuration of the VAD module. *-Dense means dense layer for the corresponding axis.}
% \vspace{-10pt}
\label{config}
\centering
\begin{tabular}[htb]{llcl}
\toprule 
%\hline
\multicolumn{1}{c}{{Layer Name}} & \multicolumn{1}{c}{{Input Size}} & {\begin{tabular}[c]{@{}c@{}}Hyper Params\end{tabular}} & \multicolumn{1}{c}{{Output Size}} \\
%\hline
\midrule
F-Dense & $\text{T} \times \text{C} \times \text{F}$ & $\text{(9, 16)}$ & $\text{T} \times \text{C} \times \text{16}$ \\ %\hline
Reshape & $\text{T} \times \text{C} \times \text{16}$ & - & $\text{T} \times \text{4C} \times \text{4}$ \\ %\hline
Maxpool1d & $\text{T} \times \text{4C} \times \text{4}$ & kernel=4, stride=4 & $\text{T} \times \text{4C} \times \text{1}$ \\ %\hline
Reshape & $\text{T} \times \text{4C} \times \text{1}$ & - & $\text{T} \times \text{C} \times \text{4}$ \\ %\hline
F-LSTM & $\text{T} \times \text{C} \times \text{4}$ & hidden size=C & $\text{T} \times \text{C} \times \text{4}$ \\ %\hline
F-Dense & $\text{T} \times \text{C} \times \text{16}$ & $\text{(16,1)}$ & $\text{T} \times \text{C} \times \text{1}$ \\ %\hline
C-Dense & $\text{T} \times \text{C} \times \text{1}$ & $\text{(C,2)}$ & $ \text{T} \times \text{2}$ \\ 
% \midrule
\bottomrule
\end{tabular}
\vspace{-10pt}
\end{table}

Inspired by~\cite{birnbaum2019temporal}, we use the norm-layer implemented by LSTM as our VAD module. But different from~\cite{birnbaum2019temporal},  our structure recurs at channel axis, leading to a causal VAD module. The specific details of the VAD module can be found in Fig.~\ref{fig:system}(d) and Table~\ref{config}, where $\otimes$ means the point-wise product of tensor $G \in \mathbb{R}^{T \times 1 \times 4 \times C}$ and tensor $H \in \mathbb{R}^{T \times 4 \times 4 \times C}$. The F-LSTM follows the details defined in our previous work~\cite{zhang2021ft}.

\subsection{Loss function}
\label{ssec:cost}

Our loss function is based on the power-law compressed phase-aware (PLCPA) loss~\cite{eskimez2021human} and the VAD information. As described in~\cite{eskimez2021human}, PLCPA is beneficial for both ASR accuracy and perceptual quality. It is mainly composed of two parts -- amplitude loss $\mathcal{L}_{\text{mag}}$ and phase loss $\mathcal{L}_{\text{pha}}$, defined as follows:
{
\setlength\abovedisplayskip{2.0pt}
\setlength\belowdisplayskip{2.0pt}
\begin{equation}
\footnotesize
\begin{aligned}
\mathcal{L}_{\text {mag}}(t, f) &= \lvert \lvert S(t, f)\rvert^{p}-\rvert \hat{S}(t, f)\rvert^{p}\rvert^{2} \\
\mathcal{L}_{\text {pha}}(t, f) &=\lvert\lvert S(t, f)\rvert^{p} e^{j \varphi(S(t, f))}-\lvert\hat{S}(t, f)\rvert^{p} e^{j \varphi(\hat{S}(t, f))}\lvert^{2}.
\end{aligned}
\end{equation}
}
The VAD loss $\mathcal{L}_{\text {vad}}$ is defined as
{
\setlength\abovedisplayskip{2.0pt}
\setlength\belowdisplayskip{2.0pt}
\begin{equation}
\footnotesize
\mathcal{L}_{\text{vad}} =\mathrm {CrossEntropy}\left({P},\bar{P} \right)
\end{equation}
}
where $\bar{P} \in \mathbb{R}^{T \times 1}$ is the near-end speech activity label based on the short-term energy threshold and $P \in \mathbb{R}^{T \times 2}$ is the estimated state of the VAD module.

\textbf{Echo-aware loss}: Previous studies~\cite{fu2021uformer} have shown that the use of multiple losses, such as SI-SNR and MSE, are beneficial, where the weight of each loss is adjusted through a hyper-parameter. Inspired by this, in our approach, we adjust the weight of different TF-bins using the proportion of echo power. So we get the echo weighted coefficients defined as
{
\setlength\abovedisplayskip{2.0pt}
\setlength\belowdisplayskip{2.0pt}
\begin{equation}
\footnotesize
\mathcal{W}_{\text{echo}}(t, f) =\frac{|Z(t, f)|^{2}}{|Z(t, f)|^{2}+|S(t, f)|^{2}}.
\end{equation}
}
Weighting the amplitude loss $\mathcal{L}_{\text {mag}}$, we get
{
\setlength\abovedisplayskip{2.0pt}
\setlength\belowdisplayskip{2.0pt}
\begin{equation}
\footnotesize
\mathcal{L}_{\text{echo}} = \frac{1}{T}\frac{1}{F}\sum_{T}\sum_{F} \left[\mathcal{L}_{\text {mag}}(t, f)(1 + \mathcal{W}_{\text{echo}}(t, f)) + \mathcal{L}_{\text {pha}}(t, f)\right ].
\end{equation}
}
When the echo component of a certain TF-bin is larger, the MSE weight of this TF-bin will also become larger without manual adjustment. The lower the SER, the more likely it is to suppress the echo; the higher the SER, the less the echoes are weighted.
% 这句话打算删掉，有点多余，后续实验的地方提到了
% \textbf{ This can guarantee the speech intelligibility without manual adjustment.} 

To further improve the amount of echo suppression and ensure the perceptual quality, we make use of the VAD information as follows.
{
\setlength\abovedisplayskip{2.0pt}
\setlength\belowdisplayskip{2.0pt}
\begin{equation}
\footnotesize
\begin{aligned}
\mathcal{W}_{\text {vad}}(t) &= \mathrm {GS}(P_{t}) \cdot W_{\text{fix}}^\mathrm{T}
\\
\mathcal{L}_{\text{mask}} &=\frac{1}{T}\frac{1}{F}\sum_{T}\sum_{F}\lvert\lvert S(t, f)\rvert^{p}-\lvert\hat{S}(t, f)|^{p} \cdot \mathcal{W}_{\text {vad}}(t)\rvert^{2},
\end{aligned}
\end{equation}
}
we use the $\mathrm {GS}$ (Gumbel SoftMax) as a continuous, differentiable approximation to $\arg\max$~\cite{jang2016categorical}. The fixed weight $W_{\text{fix}} = [0, 1]^{1 \times 2}$ and $\mathcal{W}_{\text {vad}} \in \mathbb{R}^{T\times 1}$. $\mathcal{L}_{\text{mask}}$ can punish wrong judgment on VAD of a certain frame, thus reducing near-end over suppression. 

The loss function we finally used in the submitted system is
{
\setlength\abovedisplayskip{2.0pt}
\setlength\belowdisplayskip{2.0pt}
\begin{equation}
\footnotesize
\mathcal{L}_{\text {final}} =\mathcal{L}_{\text{echo}} +0.2 \cdot  \mathcal{L}_{\text{mask}}+0.1 \cdot \mathcal{L}_{\text {vad}}.
\end{equation}
}

\begin{table*}[!htbp]
\footnotesize
\centering
\vspace{-10pt}
\setlength{\tabcolsep}{5pt}
\caption{Echo suppression performance. DT: double talk, ST: single-talk, NE: near-end, FE: far-end. WB-PESQ used for DT and ST-NE scenarios and ERLE used for ST-FE scenario in both simulated test set (ST-FE) and ICASSP 2022 blind test set (Blind). }
\label{metric1}
\begin{tabular}{ccccccccccccccccc}
\toprule
 Method & TDE & Signal & Loss & \multicolumn{6}{c}{DT}  & \multicolumn{2}{c}{ST-NE} &  \multicolumn{3}{c}{ST-FE}  & \multicolumn{1}{c}{Blind} & Data\\
\midrule
 \multicolumn{4}{c}{SNR (in dB)}&
\multicolumn{3}{c}{5} &
\multicolumn{3}{c}{+$\infty$} &
5 &
+$\infty$ &
\multicolumn{3}{c}{+$\infty$} &
 & \\
\hline
% \midrule
\multicolumn{4}{c}{SER (in dB)} &
-5 &
5 &
15  &
-5  &
5  &
15  &
\multicolumn{2}{c}{+$\infty$} &
-5  &
5  &
15  &
 & \\
% \midrule
\hline
Input&-&-&-&1.42&1.78&1.99&1.58&2.24&2.87&2.10&4.50&\multicolumn{4}{c}{0}&- \cr 
GCCRN&$\times$&$DX$&$\mathcal{L}_{\text{cmse}}$&1.83&2.25&2.37&2.15&2.79&3.18&2.48&3.55&30.22&30.17&22.47&21.76 \cr
GCCRN&$\times$&$EX$-M&$\mathcal{L}_{\text{cmse}}$&2.06&2.43&2.58&2.37&2.96&3.36&2.66&3.71&36.03&35.15&28.61&27.25 \cr
GFTNN&$\times$&$EX$-M&$\mathcal{L}_{\text{plcpa}}$&2.12&2.51&2.66&2.45&3.07&3.46&2.72&3.75&44.63&43.13&39.62&32.57 \cr
GFTNN&$\times$&$EX$-M&$\mathcal{L}_{\text{echo}}$&2.20&2.59&2.74&2.57&3.22&3.61&2.81&3.77&82.38&76.46&67.30&58.63 & 460h\cr
GFTNN&$\times$&$DEY$-M&$\mathcal{L}_{\text{echo}}$&2.27&2.66&2.83&2.62&3.24&3.65&2.91&4.14&96.69&97.52&90.37&80.45 \cr
GFTNN&$\checkmark$&$DEY$-M&$\mathcal{L}_{\text{echo}}$&2.32&2.67&2.82&2.72&3.30&3.65&2.93&4.35&95.50&95.16&89.29&76.77 \cr
GFTNN&$\checkmark$&$DEY$-W&$\mathcal{L}_{\text{echo}}$&2.38&2.74&2.85&2.74&3.32&3.70&2.99&4.35&78.24&71.33&64.25&62.32 \cr
\hline
GFTNN& & &$\mathcal{L}_{\text{echo}}$&2.44&2.80&2.96&2.79&3.36&3.73&3.04&4.35&82.36&76.71&67.69&67.98 \cr
GFTNN-VAD &$\checkmark$ & $DEY$-W &$\mathcal{L}_{\text{final}}$&2.43&2.78&2.94&2.78&3.34&3.72&3.01&4.33&81.37&76.28&65.80&70.86& 860h \cr
GFTNN-VAD-L & & &$\mathcal{L}_{\text{final}}$&2.44&2.78&2.94&2.81&3.37&3.73&3.02&4.34&84.96&83.85&77.63&79.28 \cr
\bottomrule
\end{tabular}
 \vspace{-10pt}
\end{table*}

% \vspace{-5pt}
\section{Experiments}
\label{sec:exp}
\subsection{Dataset}
\label{ssec:dataset}

In our experiments, \textit{train-clean-100} and \textit{train-clean-360} from Librispeech~\cite{panayotov2015librispeech}, which has a sampling rate of 16k Hz, together with the speech data from DNS-Challenge~\cite{reddy2021interspeech}, which has a sampling rate of both 16k and 48k Hz, are used as near-end signal and reference signal. For the noise signal, we use the noise data from DNS-Challenge, which has a sampling rate of both 16k and 48k Hz. For the echo signal, we use all the synthetic echo signals and real far-end single-talk recordings provided by the AEC-Challenge, which covers a variety of voice devices and echo signal delay. Furthermore, we also use the speech from Librispeech and DNS dataset to simulate echo data by convolving with simulated room impulse response (RIR). We use the HYB method described in~\cite{bezzam2020study} instead of the image method to generate 20,000 RIRs. The data generation scheme and the specific configuration such as room size of the RIRs are described in~\cite{zhang2021ft}.

The training set has 860 h data in total, which contains 460 h data with simulated echo and 400 h data with real-recorded echo. Development and test set share the same generating method with training set, which contain 30 h and 15 h data, respectively. The source data among these three sets has no overlap. 

\subsection{Performance metrics}
\label{ssec:metric2}
Echo suppression performance is evaluated in terms of echo return loss enhancement (ERLE) (defined in Eq.~(10)) and wide-band perceptual evaluation of speech quality (WB-PESQ)~\cite{rix2001perceptual} for single-talk (ST) periods and double-talk (DT) periods, respectively. The AEC-Challenge also provides mean opinion score (MOS) results based on the ITU-T Recommendations P.831, P.832 and P.808~\cite{cutler2021crowdsourcing}. The official ranking is based on the score which reflects both MOS and speech recognition WAcc~\cite{cutler2022AEC}.
{
\setlength\abovedisplayskip{2.0pt}
\setlength\belowdisplayskip{2.0pt}
\begin{equation}
\footnotesize
\mathrm{ERLE}=10\log_{10}\left[\sum_{n} d_{full}^{2}(n) / \sum_{n} \hat{s}_{full}^{2}(n)\right]
\end{equation}
}
\begin{table}[]
\vspace{-5pt}
\setlength{\tabcolsep}{2pt}
\footnotesize
\caption{Subjective ratings in terms of MOS and WAcc for the blind test set of the AEC-challenge. The confidence interval is 0.02. DT-Echo DMOS: echo annoyance DMOS for DT scenario. DT-Other DMOS: other impairments DMOS of DT scenario~\cite{cutler2021crowdsourcing}.}% \vspace{-1pt}
\label{metric2}
\centering

\begin{tabular}[htb]{lcccccc}
\toprule
\multicolumn{1}{c}{Method} & \begin{tabular}[c]{@{}c@{}}ST-NE\\ MOS\end{tabular} & \begin{tabular}[c]{@{}c@{}}ST-FE\\ DMOS\end{tabular} & \begin{tabular}[c]{@{}c@{}}DT-ECHO\\ DMOS\end{tabular} & \begin{tabular}[c]{@{}c@{}}DT-Other\\ DMOS\end{tabular}  &
\begin{tabular}[c]{@{}c@{}}WAcc \end{tabular} & \begin{tabular}[c]{@{}c@{}}Score\end{tabular} \\
\midrule
% \hline
Baseline~\cite{cutler2022AEC}  & 4.152  & 4.563 & 4.122 & 3.563 & 0.659 & 0.752\\
GFTNN-VAD-L  & \textbf{4.238}     & \textbf{4.801} & \textbf{4.711}  & \textbf{4.257}  & \textbf{0.817} & \textbf{0.864}\cr
\bottomrule
% \hline
\end{tabular}
\vspace{-10pt}
\end{table}

\subsection{Experimental setup}
\label{ssec:expsettings}

Window length and hop size are 20 ms and 10 ms, respectively, resulting in 30 ms overall algorithmic latency. We apply 320-point STFT to each signal to produce the complex spectra. Chunk size of our generated data is set to 10 s. Hyper-parameter $p$ is set to 0.5 for $\mathcal{L}_{\text {mag}}$ and $\mathcal{L}_{\text {pha}}$. We compare different models with various configurations, as shown in Table~\ref{metric1}. Here $\mathcal{L}_{\text{cmse}}$ refers to the loss function used in~\cite{peng2021acoustic} and $\mathcal{L}_{\text{plcpa}}=\mathcal{L}_{\text{mag}} + \mathcal{L}_{\text{pha}}$. $DX$ means we directly feed the microphone signal and the near-end speech signal into the network, without using any DSP-based linear AEC. *-M/W means using MDF/wRLS as the adaptive filter respectively. All neural models are trained with the Adam optimizer~\cite{kingma2014adam} for 60 epochs with an initial learning rate of 1e-4, and the learning rate is halved if there is no loss decrease on development set for 2 epochs. The convolution kernels move with a stride of (1, 2). The number of output channels for each layer in encoder/decoder is 80 for GFTNN and GFTNN-VAD, and 128 for GFTNN-VAD-L. The whole size of the submitted model (GFTNN-VAD-L) is 4.7912M in parameters.  Tested on Intel(R) Xeon(R) Platinum 8163 CPU@2.50GHz quad-core machine, the total real time factor (RTF) of our submitted model is 0.1706, which is composed of 0.1306 for the GFTNN post-filter and 0.04 for the DSP part, including sub-band decomposition, adaptive filter and synthesize.
Some of the processed audio clips can be found in our demo page\footnote{https://echocatzh.github.io/GFTNN}.

\subsection{Results and analysis}
\label{ssec:performace}

From Table~\ref{metric1}, we first see that the use of an adaptive filter (linear AEC) is beneficial -- directly using a neural network for AEC leads to inferior performance (comparing $DX$ with $EX$/$DEY$). We also notice that replacing $\mathcal{L}_{\text{plcpa}}$ with $\mathcal{L}_{\text{echo}}$ improves both PESQ and ERLE a lot, which proves that weighting the MSE by echo is very beneficial. Consistent with~\cite{wang2021nn3a}, replacing the input signals $EX$ with $DEY$ leads to improved PESQ and ERLE in each scenario.  In real recordings of ST-NE scenario~\cite{cutler2022AEC}, the reference signal $x(n)$ may be completely irrelevant to microphone signal $d(n)$. This means that the performance of the neural AEC may be affected if the input signals contains $X$ (such as $DX$ and $EX$), because unlike the DSP methods, it is difficult for a neural network to find consistent rules for such uncorrelated input combinations. This also explains why the input combination $DEY$ has better performance than the other two combinations. 
The distortion in the subband decomposition and synthesis process makes the PESQ upper limit to be 4.35. When the SER $=+\infty$ and SNR $=+\infty$, which means the clean signal, the use of the TDE module in the GFTNN post-filter improves the perceptual quality of speech (PESQ $>=4.3$). 
% We notice that in our linear AEC + poster-filter framework, wRLS leads to better performance than MDF (comparing \textit{DEY}-W with \textit{DEY}-M).
Comparing the two linear AEC algorithms (\textit{DEY}-W vs. \textit{DEY}-M) in our two-stage framework, wRLS leads to better performance than MDF due to its better convergence.
Finally, the larger model GFTNN-VAD-L shows generally superior performance. We process the blind test clips using this model and submit to the challenge.

There are three observations we would like to explain further. First, replacing $DEY$-M with $DEY$-W results in almost 20 dB decline in ERLE despite of improvement on PESQ. In fact, there is no obvious difference in subjective listening performance when ERLE is greater than 60 dB. Second, the VAD module does not lead to better PESQ, but we find that many emotional clips in the blind test set are not over-suppressed because of the use of the VAD module. The better perceptual listening performance on these clips indicates that the network can effectively learn to distinguish speech frame from non-speech frame. Finally, we find that when $x(n)$ is uncorrelated with $d(n)$, the complex-valued neural network will cause distortion to the estimated $\hat s(n)$. Hence in this paper, we no longer use complex-valued structure.

The officially released results on the blind test set in Table~\ref{metric2} show that our method significantly outperforms the challenge baseline with absolute 0.158 \textit{WAcc} gain and 0.112 \textit{score} gain. In other words, our system ranked the 1st place in WAcc and the 3rd place in mean opinon score (MOS) and the final \textit{score}.

% \vspace{-5pt}
\section{Conclusions}
\label{sec:end}
This paper introduces our linear AEC + neural post-filter system submitted to the ICASSP 2022 AEC challenge. With the help of the specifically designed network structure GFTNN, multi-task learning with VAD and echo-aware loss function, our proposed AEC system can achieve better echo cancellation and noise suppression performance while ensuring that the near-end speech is not over suppressed. Our submitted system manages to rank the 3rd place in the challenge with good subjective quality (MOS) and speech recognition accuracy (WAcc). In future work, we will explore whether our echo-aware weighting loss can be transferred to related tasks like target speaker extraction (TSE), and explore more cascade schemes.

% References should be produced using the bibtex program from suitable
% BiBTeX files (here: strings, refs, manuals). The IEEEbib.bst bibliography
% style file from IEEE produces unsorted bibliography list.
% -------------------------------------------------------------------------
% \footnotesize
\bibliographystyle{IEEEbib}
\bibliography{refs}

\end{document}